\documentclass[sigconf,screen]{acmart}

\AtBeginDocument{%
  \providecommand\BibTeX{{%
    Bib\TeX}}}

\setcopyright{none}

\usepackage[utf8]{inputenc}
\usepackage[T1]{fontenc}

\usepackage{pifont}
\usepackage{algorithm, algpseudocode}
\usepackage{amsmath,amsfonts}
\usepackage{graphicx}
\usepackage{textcomp}
\usepackage{tabularx}
\usepackage{color}
\usepackage{xcolor}
\usepackage{tcolorbox}
\usepackage{colortbl}
\usepackage{fvextra}
\usepackage{subcaption}
\usepackage{xspace}
\usepackage{longtable}

\usepackage{multicol}

\usepackage{enumitem}
\setlist{itemsep=1pt,topsep=1pt,parsep=0pt,partopsep=0pt}

\usepackage{multirow}
\usepackage{soul}

\definecolor{templateyellow}{RGB}{255, 250, 205}
\definecolor{templateblue}{RGB}{0, 0, 135}
\definecolor{templategrey}{RGB}{204, 204, 204}

\algrenewcommand\algorithmicindent{1.5em}
\algtext*{EndWhile}
\algtext*{EndIf}
\algtext*{EndFor}

\let\oldReturn\Return
\renewcommand{\Return}{\State\oldReturn}

\makeatletter
\newlength{\trianglerightwidth}
\algnewcommand{\LineComment}[1]{\Statex \hskip\ALG@thistlm $\triangleright$ #1}
\algnewcommand{\IndentLineComment}[1]{\Statex \hskip\ALG@tlm \(\triangleright\) #1}
\algnewcommand{\LineCommentCont}[1]{\Statex \hskip\ALG@thistlm%
  \parbox[t]{\dimexpr\linewidth-\ALG@thistlm}{\hangindent=\trianglerightwidth \hangafter=1 \strut$\triangleright$ #1\strut}}
\algnewcommand{\IndentLineCommentCont}[1]{\Statex \hskip\ALG@tlm%
  \parbox[t]{\dimexpr\linewidth-\ALG@tlm}{\hangindent=\trianglerightwidth \hangafter=1 \strut$\triangleright$ #1\strut}}
\makeatother

\def\BibTeX{{\rm B\kern-.05em{\sc i\kern-.025em b}\kern-.08em
    T\kern-.1667em\lower.7ex\hbox{E}\kern-.125emX}}

\newcommand{\tab}{\hspace*{1em}}
\newcommand{\compactline}{\looseness=-1}

\newcommand{\revision}[1]{{#1}}

\usepackage{xltabular}

\tcbuselibrary{breakable}

\begin{document}

\title{\revision{How Effective Are They? Exploring Large Language Model Based Fuzz Driver Generation}}

\author{Cen Zhang}
\orcid{0000-0001-5603-1322}
\affiliation{%
\institution{Nanyang Technological University}
\country{Singapore}
}

\author{Yaowen Zheng}
\orcid{0000-0002-8953-0782}
\affiliation{%
\institution{Nanyang Technological University}
\country{Singapore}
}
\authornote{Yaowen Zheng is the corresponding author.}

\author{Mingqiang Bai}
\orcid{0009-0002-9197-2069}
\affiliation{%
\institution{IIE, CAS; Sch of Cyber Security, UCAS}
\city{Beijing}
\country{China}
}

\author{Yeting Li}
\orcid{0000-0003-0991-4231}
\affiliation{%
{ 
\institution{IIE, CAS; Sch of Cyber Security, UCAS}
\city{Beijing}
\country{China}
}
}

\author{Wei Ma}
\orcid{0000-0002-0044-466X}
\affiliation{%
\institution{Nanyang Technological University}
\country{Singapore}
}

\author{Xiaofei Xie}
\orcid{0000-0002-1288-6502}
\affiliation{%
\institution{Singapore Management University}
\country{Singapore}
}

\author{Yuekang Li}
\orcid{0000-0003-4382-0757}
\affiliation{%
\institution{The University of New South Wales}
\city{Sydney}
\country{Australia}
}

\author{Limin Sun}
\orcid{0000-0003-2745-7521}
\affiliation{%
\institution{IIE, CAS; Sch of Cyber Security, UCAS}
\city{Beijing}
\country{China}
}

\author{Yang Liu}
\orcid{0000-0001-7300-9215}
\affiliation{%
\institution{Nanyang Technological University}
\country{Singapore}
}
\begin{abstract}



\renewcommand{\thefootnote}{}
\footnotetext{
IIE $\rightarrow$ Institute of Information Engineering, CAS $\rightarrow$ Chinese Academy of Sciences.
Email To: \href{mailto:cen001@e.ntu.edu.sg}{cen001@e.ntu.edu.sg} \& \href{mailto:yaowen.zheng@ntu.edu.sg}{yaowen.zheng@ntu.edu.sg}.
}
\renewcommand{\thefootnote}{\arabic{footnote}} 

Fuzz drivers are essential for library API fuzzing. However, automatically generating fuzz drivers is a complex task, as it demands the creation of high-quality, correct, and robust API usage code.
An LLM-based (Large Language Model) approach for generating fuzz drivers is a promising area of research.
Unlike traditional program analysis-based generators, this text-based approach is more generalized and capable of harnessing a variety of API usage information, resulting in code that is friendly for human readers.
However, there is still a lack of understanding regarding the fundamental issues on this direction, such as its effectiveness and potential challenges.

To bridge this gap, we conducted the first in-depth study targeting the important issues of using LLMs to generate effective fuzz drivers.
Our study features a curated dataset with 86 fuzz driver generation questions from 30 widely-used C projects.
Six prompting strategies are designed and tested across five state-of-the-art LLMs with five different temperature settings.
In total, our study evaluated 736,430 generated fuzz drivers, with 0.85 billion token costs (\$8,000+ charged tokens).
Additionally, we compared the LLM-generated drivers against those utilized in industry, conducting extensive fuzzing experiments (3.75 CPU-year).
Our study uncovered that:
1) While LLM-based fuzz driver generation is a promising direction, it still encounters several obstacles towards practical applications;
2) LLMs face difficulties in generating effective fuzz drivers for APIs with intricate specifics.
Three featured design choices of prompt strategies can be beneficial: issuing repeat queries, querying with examples, and employing an iterative querying process;
3) While LLM-generated drivers can yield fuzzing outcomes that are on par with those used in the industry, there are substantial opportunities for enhancement, such as extending contained API usage, or integrating semantic oracles to facilitate logical bug detection.
\revision{
Our insights have been implemented to improve the OSS-Fuzz-Gen project, facilitating practical fuzz driver generation in industry.
}
\compactline

\end{abstract}

%




\maketitle

\section{Introduction}


Fuzz testing, aka fuzzing, has become the standard approach for discovering zero-day vulnerabilities.
Fuzz drivers are necessary components for fuzzing library APIs since fuzzing requires a directly executable target program.
Essentially, a fuzz driver is a piece of code responsible for accepting mutated input from fuzzers and executing the APIs accordingly.
An effective driver must contain a correct and robust API usage since incorrect or unsound usage can result in extensive false positive or negative fuzzing results, incurring extra manual validation efforts or testing resources waste.
Due to the high standard required, fuzz drivers are typically written by human experts, which is a labor-intensive and time-consuming process.
For instance, OSS-Fuzz~\cite{oss-fuzz}, the largest fuzzing framework for open-source projects, maintains thousands of fuzz drivers written by hundreds of contributors over the past seven years.

Generative LLMs (Large Language Models) have gained significant attention for their ability in code generation tasks~\cite{chatgpt,openai2023gpt4, roziere2023codellama, luo2023wizardcoder}.
They are language models trained on vast quantities of text and code, providing a conversational workflow where natural language based queries are posed and answered.
LLM-based fuzz driver generation is an attractive direction.
On one hand, LLMs inherently support fuzz driver generation as API usage inference is a basic scenario in LLM-based code generation.
On the other hand, LLMs are lightweight and general code generation platforms.
Existing works~\cite{fuzzgen, fudge, apicraft, winnie, intelligen, rubick, utopia}, which generate drivers by learning API usage from examples, requires program analysis on examples, while LLM-based generation can mostly work on texts.
This offers enhanced generality which facilitates not only the application on massive quantity of real-world projects but also the utilization of learning inputs in different forms.
Various sources of API usage knowledge such as documentation, error information, and code snippets can be seamlessly integrated in text form, benefiting the generation.
Moreover, LLMs can generate human-friendly code.
While some research efforts have been devoted to LLM-based code generation tasks~\cite{arxiv-adaptive-unit-test-gen,zhanglingming-llm-are-zero-shot-fuzzers,jiang2023selfplanning,pearce2022asleep,jain2022jigsaw,liu2023your,oss-fuzz-gen,prompt-fuzz}, none of them can provide a fundamental understanding on this direction.
To address this gap, we conducted an empirical study for understanding the effectiveness of zero-shot fuzz driver generation using LLMs.
Note that our primary goal is to understand the basics towards generating "\textit{more}" effective fuzz drivers, rather than generating "\textit{more effective}" fuzz drivers.
This is because creating effective drivers for more targets is a more fundamental issue than improving existing ones.
Overall, four research questions are studied:
\begin{itemize}
    \item \textbf{RQ1}\tab 
    To what extent can current LLMs generate effective fuzz drivers for software testing? 
    \item \textbf{RQ2}\tab 
    What are the primary challenges associated with generating effective fuzz drivers using LLMs?
    \item \textbf{RQ3}\tab 
    What are the effectiveness and characteristics for different prompting strategies?
    \item \textbf{RQ4}\tab 
    How do LLM-generated drivers perform comparing to those practically used in the industry?
\end{itemize}

To answer these RQs, we assembled a dataset of 86 fuzz driver generation questions collected from 30 widely-used C projects from OSS-Fuzz projects.
Each question represents a library API for which a corresponding fuzz driver is needed to conduct effective fuzz testing.
We devised six prompt strategies, taking into account three key factors: the content of the prompts, the nature of interactions between the strategies and models, and the repetition of the entire query process.
Our evaluation encompassed five state-of-the-art LLMs with five different temperature settings.
The assessed LLMs included closed-source LLMs such as gpt-4-0613~\cite{gpt-4-models}, gpt-3.5-turbo-0613~\cite{gpt-3.5-models}, and text-bison-001~\cite{palm2-text-models}, as well as open-source LLMs optimized for code generation, namely, codellama-34b-instruct~\cite{codellama-34b-instruct} and wizardcoder-15b-v1.0~\cite{wizardcoder-15b-v1.0}.
For a rigorous assessment, we developed an evaluation framework automatically validating the generated drivers based on the results of compilation and short-term fuzzing, and manually crafted checkers on API usage semantic correctness.
In total, 736,430 fuzz drivers, at the cost of 0.85 billion tokens (\$8,000+ charged tokens, 0.17/0.21 billion for gpt-4-0613/gpt-3.5-turbo-0613), were evaluated.
Besides, comparison with manually written drivers in industry on code and fuzzing metrics, \textit{e.g.}, 24-hour fuzzing experiments (3.75 CPU-year), are conducted to seek practical insights.
\compactline

The overall implications for the effectiveness of using LLM to generate fuzz drivers are two-fold.
On one hand, LLMs have demonstrated outstanding performance in evaluated configurations\footnote{A configuration stands for a combination of <Model, Prompt Strategy, Temperature>.}, suggesting a strong potential for this approach.
For instance, the optimal configuration can address 91\% questions (78/86) and all top 20 configurations can address at least half of the questions.
On the other hand, resolving a question means successful generation of at least one effective fuzz driver for the assessed API, which does not necessarily imply full practicality.
For high automation and usability, three challenges have been identified:
\ding{182} Improving the success rate to reduce generation costs.
Although most questions can be resolved by LLMs, the cost can be exceptionally high.
Typically, 71\% of questions are resolved by repeating the entire query process at least five times and 45\% require repeating the process ten times.
By enhancing their accuracy, significant financial costs for automation can be saved.
\ding{183} Ensuring semantic correctness in API usage.
Occasionally, validating the effectiveness of a generated driver requires the understanding of API usage semantics.
Failed to do so can result in ineffective fuzzing with false positive or negative results.
In our evaluation, this requirement was observed in 34\% of the assessed APIs (29/86), impeding practical application.
\ding{184} Addressing complex API dependencies.
6\% of questions (5/86) cannot be resolved by any evaluated configurations since their drivers' generation requires nontrivial preparation of the API execution contexts, which cannot be appropriately hinted by any collected usage information.
For example, some drivers require a standby network server or client to be created for interacting with the target API.
These are typical cases representing complex real-world testing requirements which deserves exploration of advanced solutions.

Prompt strategy, temperature, and model are key factors considerably affect the overall performance.
Our evaluation suggests that the dominant strategy is the one incorporating three key designs: repeatedly query, query with extended information, and iterative query.
Comparing with naive strategy, its question resolve rate soars from 10\% to 91\%.
Evaluation with lower temperature settings, especially below the threshold of 1.0, have higher performance.
This is intuitive since lower temperatures lead to more consistent and predictable outputs, which fits the goal of generating an effective fuzz driver.
Besides, the optimal temperature setting in our evaluation is 0.5.
As for models, gpt-4-0613, wizardcoder-15b-v1.0 are the best closed-source, open-source models, respectively.

Fundamentally, LLMs struggle to generate fuzz drivers which require complex API usage specifics.
We identified three beneficial designs that have distinct characteristics:
\ding{182} repeatedly queries.
When the configurations are stronger, the benefits of repetition become higher.
Besides, the benefits significantly drop after the first few rounds.
A suggested repetition value is 6.
\ding{183} query with extended information.
Adding API documentation is less helpful while adding example snippets can help significantly.
Specifically, test/example files of the target project or its variants are high-quality example sources;
\ding{184} iterative queries. 
It adds a cyclic driver fix progress after the initial query, which improves LLM's performance through its step-by-step problem-solving approach and a more thorough utilization of existing usage.
Besides, all the above designs will significantly increase the token cost. 
In comparison to OSS-Fuzz drivers, LLM-generated drivers demonstrated comparable fuzzing outcomes.
However, since LLMs tend to generate fuzz drivers with minimal API usages, significant room is still left for improving generated drivers, such as expanding API usage and incorporating semantic oracles.

\revision{
To further translate our research insights into practical values, part of our prompting strategies, including checking, categorizing, and fixing driver with runtime errors, have been implemented into OSS-Fuzz-Gen~\cite{oss-fuzz-gen}, the largest LLM-based fuzz driver generation framework operated by Google OSS-Fuzz team, facilitating the continuous fuzzing of real-world projects.
}
\compactline

In summary, our contributions are:
\begin{itemize}
    \item we conducted the first in-depth study on the effectiveness of LLM-based fuzz driver generation, which showcases the potentials and challenges of this direction;
    \item we designed and implemented six driver generation strategies.
    They are evaluated in large scale, with a systematic analysis on the effectiveness, the pros and the cons;
    \item we compared generated drivers with industrial used ones, and summarized the implications on future improvements.
    \item \revision{we ported our strategies to improve the largest industrial fuzz driver generation framework, facilitating the continuous fuzzing of hundreds open-source projects.}
\end{itemize}

\section{Preliminaries}
\label{sec:preliminaries}

\begin{figure}[t]
    \centering
    \includegraphics[width=1.0\columnwidth]{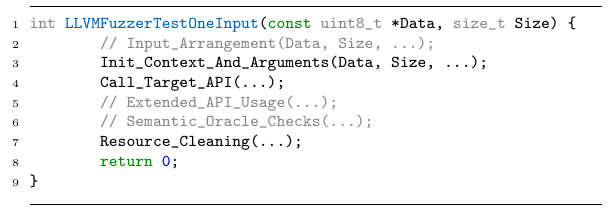}
    \caption{Key Components of A Fuzz Driver.
    }
    \label{fig:fuzz-driver-internal}
\end{figure}

\begin{figure*}[t]
    \centering
    \includegraphics[width=1.0\textwidth]{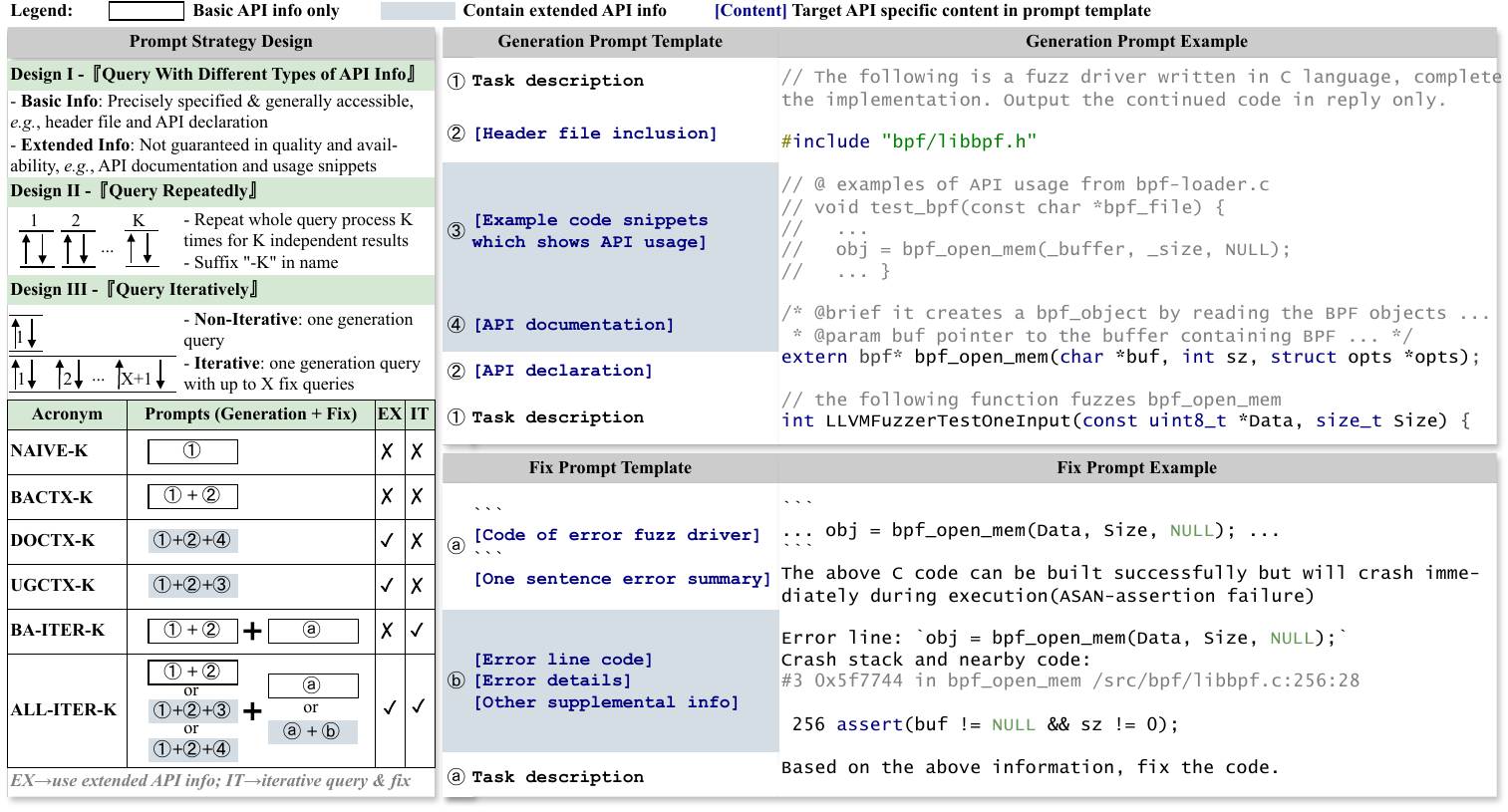}
    \caption{
        \revision{Prompt Strategies Overview.}
    \footnotesize{
        \revision{
        K is 1 or 40 in our evaluation and X is 5. The examples are simplified for demonstration purpose.
    In the fix prompt example, the driver error is caused by missing check of \textbf{Size} > 0 before calling the API, and the nearby code of \texttt{\#3} stack frame hints the error.
        }
    }
    }
    \label{fig:prompt-strategy-design}
\end{figure*}

\noindent
\textbf{Fuzz Driver Basics.} \tab 
The key components of a fuzz driver are illustrated in Figure~\ref{fig:fuzz-driver-internal}.
A typical fuzz driver has three necessary parts: prerequisites initialization (line 3), execution (line 4), and post-cleaning (line 7).
Besides, there are three optional parts commetned in lines 2, 5, and 6 that can improve a driver's effectiveness.
Line 2 part improves a driver by proper input arrangement such as rejecting too short or too long inputs, interpreting input data as multiple testing arguments, etc.
Line 5 part enables a driver to call more APIs which triggers more program behaviors during fuzzing.
Finally, line 6 part adds semantic oracles for detecting logical bugs.
These oracles are similar to \texttt{assert} statements in unit tests, aborting execution when certain program properties are unsatisfied.
Since a driver will be repeatedly executed with randomly mutated input, there is a high requirement on its correctness and robustness.
Incorrect or unrobust usage can lead to both false positives and negatives.
For instance, if a driver failed to feed the mutated data into the API, its fuzzing can never find any bug.
Or if an API argument is incorrectly initialized, false crashes may be raised.

\noindent
\textbf{Minimal Requirements of Effective Fuzz Drivers.} \tab 
The minimal requirements covers the line 3,4, and 7 of Figure~\ref{fig:fuzz-driver-internal}, which mainly include correctly initializing the arguments and satisfying necessary control flow dependencies.
Argument initialization can be one of the following cases (in the order of simplicity):
\ding{182} \textbf{C1}: If the argument value can be any value or should be naive values like \texttt{0} or \texttt{NULL}, a variable is declared or a literal constant is used directly;
\ding{183} \textbf{C2}: If the argument is supposed to be a macro or a global variable that is already defined in common libraries or the target API's project, it is located and used;
\ding{184} \textbf{C3}: If creating the argument requires the use of common library APIs, such as creating a file and writing specific content, common practices are followed;
\ding{185} \textbf{C4}: If initializing the argument requires the output of other APIs within the project, those APIs are initialized first following the above initialization cases.

\section{Methodology}
\label{sec:methodology}

\subsection{Design of Prompt Strategies}


\revision{
Figure~\ref{fig:prompt-strategy-design} illustrates our designed prompt strategies.
From left to right, the figure first provides a tabular overview for all proposed strategies, then details two types of prompt templates involved, and lastly maps the templates to concrete query examples.
Note that the listed examples are simplified for demonstration purposes, while unmodified real-world examples for each strategy can be found at \cite{fuzz-drvier-study-website}.
}
\compactline

\revision{
\noindent
\textbf{Key Designs.} \tab 
As shown in the top-left side of Figure~\ref{fig:prompt-strategy-design}, there are three key designs for prompt strategies, including query with different types of API information, query repeatedly, and query iteratively.
Design I aims for understanding the generation effectiveness given different API information as query contexts.
The information is divided as two types: the basic API information and the extended.
The former includes fundamental information such as header file name and API declaration, which are precisely specified and generally accessible in library API fuzzing scenario, while the latter requires additional resources like API-specific documentation or usage example code snippets, whose quality and availability vary for different targets.
To account for the inherent randomness in LLM output generation, design II, repeatedly query, is introduced.
Given repetition time as value K, the entire query process of a strategy will be repeated K times ($K \ge 1$), generating K independent drivers.
The maximum value of K is set as 40 in our study.
This is an empirically value we believe is comprehensive enough to understand the effectiveness of repetition.
Design III is used to understand the effectiveness of different query workflows.
The driver generation in non-iterative strategies follows a one-and-done manner where the final driver is synthesized via a single query without further refinement.
Iterative strategies have a generate-and-fix workflow.
If the driver generated in first query fails to pass the automatic validation, subsequent fix prompts are composed based on the error feedback and queried.
The fix continues until the driver passes validation or a pre-defined maximum number of iterations is reached.
The iteration is limit as five in our evaluation.
}

\revision{
\noindent
\textbf{Acronym.} \tab 
Strategies are named by concatenating the abbreviations of the three key designs.
For all strategies, there is a suffix "K" indicating that the repetition times of their query process.
If a strategy name contains "ITER", it is an iterative prompt strategy.
Otherwise, it is non-iterative.
Besides, different combinations of API information used in generation prompt have different abbreviations.
As shown in the bottom-left side of the Figure~\ref{fig:prompt-strategy-design}, there are four different combinations: the \textbf{NAIVE} query context (abbr as NAIVE, \ding{172}), the \textbf{BA}sic query \textbf{C}on\textbf{T}e\textbf{X}t (abbr as BACTX, \ding{172} + \ding{173}), extending API \textbf{DOC}umentation to basic \textbf{C}on\textbf{T}e\textbf{X}t (abbr as DOCTX, \ding{172} + \ding{173} + \ding{175}), and extending example \textbf{U}sa\textbf{G}e code snippets to the basic \textbf{C}on\textbf{T}e\textbf{X}t (abbr as UGCTX, \ding{172} + \ding{173} + \ding{174}).
Lastly, the prefix ALL in ALL-ITER-K indicates that its prompts can be the prompt designed in any other strategies.
}

\noindent
\textbf{NAIVE-K \& BACTX-K.} \tab 
\revision{
Both two strategies only use basic API information in query and non-iterative workflow.
Their only difference is the richness of the prompt context information.
}
Specifically, NAIVE-K directly asks LLMs to implement the fuzz driver solely based on a specified function name, while BACTX-K provides a basic description of the API.
In prompts of BACTX-K, it first indicates the task scope using \texttt{\#include} statement, then provides the function declaration, and finally requests implementation.
The declaration is extracted from the Abstract Syntax Tree (AST) of the header file, including both the signature and argument variable names.
\compactline

\noindent
\textbf{DOCTX-K \& UGCTX-K.} \tab 
These two strategies are extended from BACTX-K by adding extended usage information in query.
Their effectiveness represents the effects of two types of extended information: API documentation and example code snippets. 
Note that, for DOCTX-K, not all APIs have associated documentation (49/86 questions in our study).
The documentation of 20 questions was automatically extracted from the header files, while the remaining 29 were manually collected from sources like project websites, repositories, and developer manuals. 
\revision{
For UGCTX-K, example code snippets of an API are collected as follows:
\ding{182} retrieving the files containing usage code via \texttt{SourceGraph} \texttt{cli}~\cite{sourcegraph-cli-tool}.
This is a keyword search among all public code repositories including Github, Gitlab, etc.
The crawling command is \texttt{src search -json "file:.*\.c lang:c count:all {API}"} where \texttt{API} should be replaced by the target API name.
\ding{183} identifying and excluding fuzz drivers by removing the files containing function \texttt{LLVMFuzzerTestOneInput}.
\ding{184} extracting all functions directly calling the target API as example code snippets via \texttt{ANTLR} based source code analysis.
\ding{185} deduplicating the snippets by if the Jaccard Similarity~\cite{jaccard-similarity} of any two snippets $\geq$95\%.
UGCTX-K will randomly use one snippet in the prompt.
For snippet that was too long to be included into prompt, it is truncated line by line until satisfying the token length limitation.
}

\revision{
\noindent
\textbf{BA-ITER-K \& ALL-ITER-K.} \tab 
Iterative strategies have two types of prompt templates for initial generation query and subsequent fix query.
The initial generation prompt can be either of BACTX-K's, DOCTX-K's, and UGCTX-K's.
As for fix queries, we have designed seven fix templates to address seven prevalent types of errors in the generated drivers.
They follow one general fix template shown in Figure~\ref{fig:prompt-strategy-design} but are filled with error type specific details.
Due to the page limit, we discuss the key concepts of these fix prompts, leaving the detailed designs and examples in~\cite{fuzz-drvier-study-website}.
These errors are of compilation errors (1/7), linkage errors (1/7), and fuzzing runtime errors (5/7).
The error information (abbr of \texttt{\color{templateblue}[One sentence error summary]}, \texttt{\color{templateblue}[Error line code]}, and \texttt{\color{templateblue}[Error details]}) for the first two error types can be programmatically retrieved from the compiler.
According to different abnormal behaviors observed in fuzzing, fuzzing runtime errors have five subtypes, including memory leakage, out-of-memory, timeout, crash, and non-effective fuzzing (no coverage increase in one-minute short-term fuzzing).
The error information of runtime errors are retrieved by extracting the crash stacks and sanitizer summary from \texttt{libfuzzer} logs.
Lastly, for the errors that can locate its error line, we further infer its root cause API and fill API information like declaration, documentation, or usage snippets into \texttt{\color{templateblue}[Other supplemental information]} in fix prompt.
For simplicity, root cause API is identified by naively finding the last executed API based on the error line located.
}

\revision{
Note that iterative strategies exclusively utilize automated checkers, ensuring that the manually crafted semantic checkers described in Section~\ref{sec:eval-framework} are only used for thorough evaluation of the effectiveness of strategies.
The principal distinction between the two iterative strategies, BA-ITER-K and ALL-ITER-K, lies in the scope of information utilized within the queries.
BA-ITER-K confines its use to only basic API details and error information, whereas ALL-ITER-K encompasses all available information.
As shown in Figure~\ref{fig:prompt-strategy-design}, this leads to multiple options for the included extended information.
The \textit{ALL-ITER-K} strategy selects options randomly.
}

\vspace{-2pt}
\subsection{Evaluation Framework}
\label{sec:eval-framework}

\noindent
\textbf{Evaluation Question Collection.} \tab 
One question used in evaluation is simply designed as generating fuzz drivers for one given API.
However, not all APIs are suitable to be set as questions.
Naively collecting all APIs from projects will lead to the creation of meaningless or confusing questions which influences the evaluation result.
Specifically, some APIs, such as \texttt{void libxxx\_init(void)}, are meaningless fuzz targets since the code executed behind the APIs can not be affected by any input data.
Some APIs can only be supplemental APIs rather than the main fuzzing target due to the nature of their functionalities.
For example, given two APIs \texttt{object *parse\_from\_str(char *input)} and \texttt{void free\_object(object *obj)}, feeding the mutated data into \texttt{input} is apparently a better choice than feeding a meaningless pointer to the \texttt{obj} argument.
However, calling the latter API when fuzzing the former is meaningful since 
\ding{182} it may uncover the hidden bug when the former does not correctly initialize the \texttt{object *} by freeing the members of \texttt{object};
\ding{183} it releases the resources allocated in this iteration of fuzzing, which prevents the falsely reported memory leak issue.

To guarantee the collected APIs are qualified and representative, we collected the core APIs of existing fuzz drivers from OSS-Fuzz projects.
A driver's core APIs are identified using the following criteria:
\ding{182} they are the target APIs explicitly pointed out by the author in its driver file name or the code comments, \textit{e.g.}, \texttt{dns\_name\_fromwire} is the core API of driver \textit{dns\_name\_fromwire.c};
\ding{183} otherwise, we pick the basic APIs as the core rather than the supplemental ones.
For example, we picked the former between parse and use/free APIs.
For the fuzz drivers which are composite drivers fuzzing multiple APIs simultaneously, we identified multiple core APIs from them.
Specifically, we randomly selected 30 projects from OSS-Fuzz (commit \texttt{135b000926}) C projects, manually extracted 86 core APIs from 51 fuzz drivers.
Full list of questions are post at~\cite{fuzz-drvier-study-website}.

\begin{figure}[t]
    \centering
    \includegraphics[width=1.0\columnwidth]{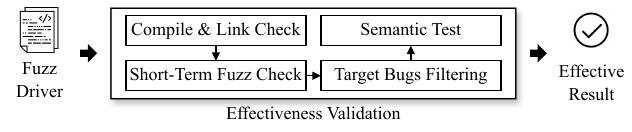}
    \caption{
        \revision{
        Fuzz Driver Effectiveness Validation Process.
        }
        }
    \label{fig:validation-checker}
\end{figure}

\revision{
\noindent
\textbf{Effectiveness Validation Criteria.} \tab 
Assessing the effectiveness of a generated fuzz driver is complex since identifying both false positives (bugs caused by the driver code) and negatives (can never find bugs given its incorrect usage) rely on the understanding of API usage semantics.
Figure~\ref{fig:validation-checker} is a streamlined four-step semi-automatic validation process:
\ding{182} Use a compiler to check for grammatical errors in the driver code.
\ding{183} Observe the driver in a one-minute fuzzing session starting with no initial seed.
It is ineffective if it either fails to show coverage progress or reports any bugs.
The assumption behind is that, given a poor fuzzing setup, neither the zero coverage progress nor the quick identification of bugs for a well-tested API are normal behaviours.
Considering that this criteria can still lead to incorrect validation, two additional steps are introduced for result refinements.
\ding{184} If the driver reports bugs in \ding{183}, we filter the true bugs contained inside.
This is done by first collecting target true bugs via ten cpu-day fuzzing using existing OSS-Fuzz drivers, then manually building filters based on the root causes of these bugs.
\ding{185} For drivers reporting no bugs after \ding{184}, we check whether they are substantially testing the target API or not.
To this end, we write API-specific semantic tests to detect common ineffective patterns observed in LLM-generated fuzz drivers.
The tests include verifying the target API is called for all 86 questions, checking the correct usage of fuzzing data to key arguments for 8 questions (such as fuzzing file contents instead of file names), ensuring critical dependent APIs are invoked in 16 questions, and confirming necessary execution contexts are prepared for 5 questions (for instance, having a standby server process available for testing client APIs).
We implement these tests by injecting hooking code into the driver.
For more details on these tests, please refer to our website~\cite{fuzz-drvier-study-website}.
}
\compactline

\noindent
\textbf{Evaluation Configuration.} \tab 
In this paper, a configuration represents a specific combination of the three factors: the LLM used, the prompt strategy employed, and the selected temperature setting, abbr as <model, prompt strategy, temperature>.
As shown in Table~\ref{tab:overall_eva_rslt}, we evaluated six prompt strategies on five LLMs with five different temperatures.
A configuration's top\_p is set as its model's default value.
And the system role~\cite{system-role-usage} is set as "\sethlcolor{templategrey}\hl{\texttt{You are a security auditor who writes fuzz drivers for library APIs.}}".


\section{Overall Effectiveness (\textbf{RQ1})}





\begin{table}[!t]
\centering
\caption{Overall Evaluation Result.
\footnotesize{K represents 40, "-" means failed to retrieve full query results or not applicable for the given model.}
}
\label{tab:overall_eva_rslt}
\resizebox{1.0\linewidth}{!}{
\setlength\arrayrulewidth{0.1pt}
\begin{tabular}{cl!{\color{white}\vrule}c!{\color{white}\vrule}c!{\color{white}\vrule}c!{\color{white}\vrule}c!{\color{white}\vrule}c}
& & \multicolumn{5}{c}{Temperature} \tabularnewline
\multicolumn{2}{c}{Strategy, Model} & \cellcolor{black!20}\texttt{0.0} & \cellcolor{black!20}\texttt{0.5} & \cellcolor{black!20}\texttt{1.0} & \cellcolor{black!20}\texttt{1.5} & \cellcolor{black!20}\texttt{2.0} \tabularnewline
\arrayrulecolor{white}\hline\hline
\cellcolor{black!0}& \cellcolor{black!20} {\normalsize gpt-4-0613}& \cellcolor{green!10}{\large 9}/{\footnotesize 86}& \cellcolor{green!10}{\large 9}/{\footnotesize 86}& \cellcolor{green!10}{\large 9}/{\footnotesize 86}& \cellcolor{green!0}{\large 0}/{\footnotesize 86}& {\large -}{\tiny -} \tabularnewline\arrayrulecolor{white}\hline
\cellcolor{black!0}& \cellcolor{black!20} {\normalsize gpt-3.5-turbo-0613}& \cellcolor{green!0}{\large 0}/{\footnotesize 86}& \cellcolor{green!10}{\large 1}/{\footnotesize 86}& \cellcolor{green!0}{\large 0}/{\footnotesize 86}& \cellcolor{green!0}{\large 0}/{\footnotesize 86}& \cellcolor{green!0}{\large 0}/{\footnotesize 86} \tabularnewline\arrayrulecolor{white}\hline
\cellcolor{black!0}& \cellcolor{black!20} {\normalsize wizardcoder-15b-v1.0}& \cellcolor{green!10}{\large 3}/{\footnotesize 86}& \cellcolor{green!10}{\large 1}/{\footnotesize 86}& \cellcolor{green!10}{\large 1}/{\footnotesize 86}& \cellcolor{green!0}{\large 0}/{\footnotesize 86}& \cellcolor{green!0}{\large 0}/{\footnotesize 86} \tabularnewline\arrayrulecolor{white}\hline
\cellcolor{black!0}& \cellcolor{black!20} {\normalsize text-bison-001}& \cellcolor{green!10}{\large 2}/{\footnotesize 86}& \cellcolor{green!10}{\large 2}/{\footnotesize 86}& \cellcolor{green!10}{\large 1}/{\footnotesize 86}& {\large -}{\tiny -}& {\large -}{\tiny -} \tabularnewline\arrayrulecolor{white}\hline
\multirow{-5}{*}{\cellcolor{black!0} \rotatebox[origin=c]{90}{{NAIVE-1}}}& \cellcolor{black!20} {\normalsize codellama-34b-instruct}& \cellcolor{green!0}{\large 0}/{\footnotesize 86}& \cellcolor{green!0}{\large 0}/{\footnotesize 86}& \cellcolor{green!0}{\large 0}/{\footnotesize 86}& \cellcolor{green!0}{\large 0}/{\footnotesize 86}& \cellcolor{green!0}{\large 0}/{\footnotesize 86} \tabularnewline\arrayrulecolor{white}\hline
\arrayrulecolor{white}\hline\hline
\cellcolor{black!0}& \cellcolor{black!20} {\normalsize gpt-4-0613}& \cellcolor{green!20}{\large 12}/{\footnotesize 86}& \cellcolor{green!40}{\large 30}/{\footnotesize 86}& \cellcolor{green!40}{\large 30}/{\footnotesize 86}& \cellcolor{green!10}{\large 5}/{\footnotesize 86}& {\large -}{\tiny -} \tabularnewline\arrayrulecolor{white}\hline
\cellcolor{black!0}& \cellcolor{black!20} {\normalsize gpt-3.5-turbo-0613}& \cellcolor{green!0}{\large 0}/{\footnotesize 86}& \cellcolor{green!10}{\large 6}/{\footnotesize 86}& \cellcolor{green!10}{\large 8}/{\footnotesize 86}& \cellcolor{green!10}{\large 8}/{\footnotesize 86}& \cellcolor{green!0}{\large 0}/{\footnotesize 86} \tabularnewline\arrayrulecolor{white}\hline
\cellcolor{black!0}& \cellcolor{black!20} {\normalsize wizardcoder-15b-v1.0}& \cellcolor{green!10}{\large 3}/{\footnotesize 86}& \cellcolor{green!10}{\large 8}/{\footnotesize 86}& \cellcolor{green!20}{\large 11}/{\footnotesize 86}& \cellcolor{green!10}{\large 1}/{\footnotesize 86}& \cellcolor{green!0}{\large 0}/{\footnotesize 86} \tabularnewline\arrayrulecolor{white}\hline
\cellcolor{black!0}& \cellcolor{black!20} {\normalsize text-bison-001}& \cellcolor{green!10}{\large 2}/{\footnotesize 86}& \cellcolor{green!10}{\large 5}/{\footnotesize 86}& \cellcolor{green!10}{\large 5}/{\footnotesize 86}& {\large -}{\tiny -}& {\large -}{\tiny -} \tabularnewline\arrayrulecolor{white}\hline
\multirow{-5}{*}{\cellcolor{black!0} \rotatebox[origin=c]{90}{{NAIVE-K}}}& \cellcolor{black!20} {\normalsize codellama-34b-instruct}& \cellcolor{green!0}{\large 0}/{\footnotesize 86}& \cellcolor{green!10}{\large 1}/{\footnotesize 86}& \cellcolor{green!10}{\large 3}/{\footnotesize 86}& \cellcolor{green!0}{\large 0}/{\footnotesize 86}& \cellcolor{green!0}{\large 0}/{\footnotesize 86} \tabularnewline\arrayrulecolor{white}\hline
\arrayrulecolor{white}\hline\hline
\cellcolor{black!0}& \cellcolor{black!20} {\normalsize gpt-4-0613}& \cellcolor{green!40}{\large 29}/{\footnotesize 86}& \cellcolor{green!50}{\large 41}/{\footnotesize 86}& \cellcolor{green!50}{\large 41}/{\footnotesize 86}& \cellcolor{green!30}{\large 21}/{\footnotesize 86}& {\large -}{\tiny -} \tabularnewline\arrayrulecolor{white}\hline
\cellcolor{black!0}& \cellcolor{black!20} {\normalsize gpt-3.5-turbo-0613}& \cellcolor{green!20}{\large 12}/{\footnotesize 86}& \cellcolor{green!40}{\large 29}/{\footnotesize 86}& \cellcolor{green!40}{\large 30}/{\footnotesize 86}& \cellcolor{green!30}{\large 24}/{\footnotesize 86}& \cellcolor{green!10}{\large 1}/{\footnotesize 86} \tabularnewline\arrayrulecolor{white}\hline
\cellcolor{black!0}& \cellcolor{black!20} {\normalsize wizardcoder-15b-v1.0}& \cellcolor{green!10}{\large 7}/{\footnotesize 86}& \cellcolor{green!30}{\large 23}/{\footnotesize 86}& \cellcolor{green!30}{\large 25}/{\footnotesize 86}& \cellcolor{green!20}{\large 17}/{\footnotesize 86}& \cellcolor{green!0}{\large 0}/{\footnotesize 86} \tabularnewline\arrayrulecolor{white}\hline
\cellcolor{black!0}& \cellcolor{black!20} {\normalsize text-bison-001}& \cellcolor{green!10}{\large 7}/{\footnotesize 86}& \cellcolor{green!20}{\large 13}/{\footnotesize 86}& \cellcolor{green!20}{\large 15}/{\footnotesize 86}& {\large -}{\tiny -}& {\large -}{\tiny -} \tabularnewline\arrayrulecolor{white}\hline
\multirow{-5}{*}{\cellcolor{black!0} \rotatebox[origin=c]{90}{{BACTX-K}}}& \cellcolor{black!20} {\normalsize codellama-34b-instruct}& \cellcolor{green!0}{\large 0}/{\footnotesize 86}& \cellcolor{green!10}{\large 1}/{\footnotesize 86}& \cellcolor{green!20}{\large 11}/{\footnotesize 86}& \cellcolor{green!0}{\large 0}/{\footnotesize 86}& \cellcolor{green!0}{\large 0}/{\footnotesize 86} \tabularnewline\arrayrulecolor{white}\hline
\arrayrulecolor{white}\hline\hline
\cellcolor{black!0}& \cellcolor{black!20} {\normalsize gpt-4-0613}& \cellcolor{green!40}{\large 29}/{\footnotesize 86}& \cellcolor{green!50}{\large 40}/{\footnotesize 86}& \cellcolor{green!50}{\large 41}/{\footnotesize 86}& \cellcolor{green!30}{\large 22}/{\footnotesize 86}& {\large -}{\tiny -} \tabularnewline\arrayrulecolor{white}\hline
\cellcolor{black!0}& \cellcolor{black!20} {\normalsize gpt-3.5-turbo-0613}& \cellcolor{green!20}{\large 11}/{\footnotesize 86}& \cellcolor{green!30}{\large 22}/{\footnotesize 86}& \cellcolor{green!40}{\large 29}/{\footnotesize 86}& \cellcolor{green!30}{\large 24}/{\footnotesize 86}& \cellcolor{green!10}{\large 1}/{\footnotesize 86} \tabularnewline\arrayrulecolor{white}\hline
\cellcolor{black!0}& \cellcolor{black!20} {\normalsize wizardcoder-15b-v1.0}& \cellcolor{green!10}{\large 7}/{\footnotesize 86}& \cellcolor{green!30}{\large 24}/{\footnotesize 86}& \cellcolor{green!30}{\large 25}/{\footnotesize 86}& \cellcolor{green!20}{\large 12}/{\footnotesize 86}& \cellcolor{green!0}{\large 0}/{\footnotesize 86} \tabularnewline\arrayrulecolor{white}\hline
\cellcolor{black!0}& \cellcolor{black!20} {\normalsize text-bison-001}& \cellcolor{green!10}{\large 9}/{\footnotesize 86}& \cellcolor{green!20}{\large 14}/{\footnotesize 86}& \cellcolor{green!20}{\large 14}/{\footnotesize 86}& {\large -}{\tiny -}& {\large -}{\tiny -} \tabularnewline\arrayrulecolor{white}\hline
\multirow{-5}{*}{\cellcolor{black!0} \rotatebox[origin=c]{90}{{DOCTX-K}}}& \cellcolor{black!20} {\normalsize codellama-34b-instruct}& \cellcolor{green!0}{\large 0}/{\footnotesize 86}& \cellcolor{green!10}{\large 9}/{\footnotesize 86}& \cellcolor{green!20}{\large 13}/{\footnotesize 86}& \cellcolor{green!10}{\large 1}/{\footnotesize 86}& \cellcolor{green!0}{\large 0}/{\footnotesize 86} \tabularnewline\arrayrulecolor{white}\hline
\arrayrulecolor{white}\hline\hline
\cellcolor{black!0}& \cellcolor{black!20} {\normalsize gpt-4-0613}& \cellcolor{green!70}{\large 55}/{\footnotesize 86}& \cellcolor{green!80}{\large 63}/{\footnotesize 86}& \cellcolor{green!80}{\large 62}/{\footnotesize 86}& \cellcolor{green!30}{\large 26}/{\footnotesize 86}& {\large -}{\tiny -} \tabularnewline\arrayrulecolor{white}\hline
\cellcolor{black!0}& \cellcolor{black!20} {\normalsize gpt-3.5-turbo-0613}& \cellcolor{green!40}{\large 30}/{\footnotesize 86}& \cellcolor{green!60}{\large 47}/{\footnotesize 86}& \cellcolor{green!50}{\large 43}/{\footnotesize 86}& \cellcolor{green!40}{\large 31}/{\footnotesize 86}& \cellcolor{green!0}{\large 0}/{\footnotesize 86} \tabularnewline\arrayrulecolor{white}\hline
\cellcolor{black!0}& \cellcolor{black!20} {\normalsize wizardcoder-15b-v1.0}& \cellcolor{green!50}{\large 39}/{\footnotesize 86}& \cellcolor{green!60}{\large 50}/{\footnotesize 86}& \cellcolor{green!60}{\large 48}/{\footnotesize 86}& \cellcolor{green!20}{\large 13}/{\footnotesize 86}& \cellcolor{green!0}{\large 0}/{\footnotesize 86} \tabularnewline\arrayrulecolor{white}\hline
\cellcolor{black!0}& \cellcolor{black!20} {\normalsize text-bison-001}& \cellcolor{green!30}{\large 21}/{\footnotesize 86}& \cellcolor{green!40}{\large 27}/{\footnotesize 86}& \cellcolor{green!50}{\large 38}/{\footnotesize 86}& {\large -}{\tiny -}& {\large -}{\tiny -} \tabularnewline\arrayrulecolor{white}\hline
\multirow{-5}{*}{\cellcolor{black!0} \rotatebox[origin=c]{90}{{UGCTX-K}}}& \cellcolor{black!20} {\normalsize codellama-34b-instruct}& \cellcolor{green!0}{\large 0}/{\footnotesize 86}& \cellcolor{green!10}{\large 8}/{\footnotesize 86}& \cellcolor{green!30}{\large 21}/{\footnotesize 86}& \cellcolor{green!0}{\large 0}/{\footnotesize 86}& \cellcolor{green!0}{\large 0}/{\footnotesize 86} \tabularnewline\arrayrulecolor{white}\hline
\arrayrulecolor{white}\hline\hline
\cellcolor{black!0}& \cellcolor{black!20} {\normalsize gpt-4-0613}& \cellcolor{green!70}{\large 56}/{\footnotesize 86}& \cellcolor{green!70}{\large 57}/{\footnotesize 86}& \cellcolor{green!80}{\large 62}/{\footnotesize 86}& \cellcolor{green!30}{\large 23}/{\footnotesize 86}& {\large -}{\tiny -} \tabularnewline\arrayrulecolor{white}\hline
\cellcolor{black!0}& \cellcolor{black!20} {\normalsize gpt-3.5-turbo-0613}& \cellcolor{green!40}{\large 32}/{\footnotesize 86}& \cellcolor{green!60}{\large 47}/{\footnotesize 86}& \cellcolor{green!50}{\large 43}/{\footnotesize 86}& \cellcolor{green!40}{\large 28}/{\footnotesize 86}& \cellcolor{green!10}{\large 2}/{\footnotesize 86} \tabularnewline\arrayrulecolor{white}\hline
\cellcolor{black!0}& \cellcolor{black!20} {\normalsize wizardcoder-15b-v1.0}& \cellcolor{green!10}{\large 8}/{\footnotesize 86}& \cellcolor{green!30}{\large 24}/{\footnotesize 86}& \cellcolor{green!50}{\large 37}/{\footnotesize 86}& \cellcolor{green!20}{\large 13}/{\footnotesize 86}& \cellcolor{green!0}{\large 0}/{\footnotesize 86} \tabularnewline\arrayrulecolor{white}\hline
\cellcolor{black!0}& \cellcolor{black!20} {\normalsize text-bison-001}& \cellcolor{green!10}{\large 9}/{\footnotesize 86}& \cellcolor{green!20}{\large 15}/{\footnotesize 86}& \cellcolor{green!30}{\large 20}/{\footnotesize 86}& {\large -}{\tiny -}& {\large -}{\tiny -} \tabularnewline\arrayrulecolor{white}\hline
\multirow{-5}{*}{\cellcolor{black!0} \rotatebox[origin=c]{90}{{BA-ITER-K}}}& \cellcolor{black!20} {\normalsize codellama-34b-instruct}& \cellcolor{green!10}{\large 6}/{\footnotesize 86}& \cellcolor{green!40}{\large 28}/{\footnotesize 86}& \cellcolor{green!30}{\large 22}/{\footnotesize 86}& \cellcolor{green!0}{\large 0}/{\footnotesize 86}& \cellcolor{green!0}{\large 0}/{\footnotesize 86} \tabularnewline\arrayrulecolor{white}\hline
\arrayrulecolor{white}\hline\hline
\cellcolor{black!0}& \cellcolor{black!20} {\normalsize gpt-4-0613}& \cellcolor{green!90}{\large 77}/{\footnotesize 86}& \cellcolor{green!90}{\large 78}/{\footnotesize 86}& \cellcolor{green!90}{\large 76}/{\footnotesize 86}& \cellcolor{green!30}{\large 25}/{\footnotesize 86}& {\large -}{\tiny -} \tabularnewline\arrayrulecolor{white}\hline
\cellcolor{black!0}& \cellcolor{black!20} {\normalsize gpt-3.5-turbo-0613}& \cellcolor{green!80}{\large 65}/{\footnotesize 86}& \cellcolor{green!80}{\large 68}/{\footnotesize 86}& \cellcolor{green!80}{\large 65}/{\footnotesize 86}& \cellcolor{green!50}{\large 37}/{\footnotesize 86}& \cellcolor{green!0}{\large 0}/{\footnotesize 86} \tabularnewline\arrayrulecolor{white}\hline
\cellcolor{black!0}& \cellcolor{black!20} {\normalsize wizardcoder-15b-v1.0}& \cellcolor{green!50}{\large 41}/{\footnotesize 86}& \cellcolor{green!60}{\large 48}/{\footnotesize 86}& \cellcolor{green!70}{\large 53}/{\footnotesize 86}& \cellcolor{green!20}{\large 11}/{\footnotesize 86}& \cellcolor{green!0}{\large 0}/{\footnotesize 86} \tabularnewline\arrayrulecolor{white}\hline
\cellcolor{black!0}& \cellcolor{black!20} {\normalsize text-bison-001}& \cellcolor{green!30}{\large 21}/{\footnotesize 86}& \cellcolor{green!50}{\large 37}/{\footnotesize 86}& \cellcolor{green!50}{\large 42}/{\footnotesize 86}& {\large -}{\tiny -}& {\large -}{\tiny -} \tabularnewline\arrayrulecolor{white}\hline
\multirow{-5}{*}{\cellcolor{black!0} \rotatebox[origin=c]{90}{{ALL-ITER-K}}}& \cellcolor{black!20} {\normalsize codellama-34b-instruct}& \cellcolor{green!20}{\large 13}/{\footnotesize 86}& \cellcolor{green!20}{\large 18}/{\footnotesize 86}& \cellcolor{green!30}{\large 26}/{\footnotesize 86}& \cellcolor{green!10}{\large 1}/{\footnotesize 86}& \cellcolor{green!0}{\large 0}/{\footnotesize 86} \tabularnewline\arrayrulecolor{white}\hline
\arrayrulecolor{white}\hline\hline

\end{tabular}
}
\end{table}

Table~\ref{tab:overall_eva_rslt} presents the results of all evaluated configurations.
The principal data displayed in the table are the question solve rates, formatted as X/Y, where X denotes the number of questions a language model successfully solves, and Y represents the total number of questions presented.
A configuration is considered to have solved a question if at least one effective fuzz driver has been generated.
For gpt-4-0613, temperature 2.0 results were incomplete due to the services' slow response time in extreme temperature settings~\cite{fuzz-drvier-study-website}.
For the text-bison-001, Google's query API limits the requests with a temperature setting above 1.0.
Nevertheless, given the poor or even zero performance of all other models with a temperatures setting 2.0, the data absence does not substantially affect our evaluation.
\revision{
This table only lists the number of solved questions while \cite{fuzz-drvier-study-website} posts full evaluation details of each model such as success rate per question.
}
\compactline

\textbf{Overall, the results offer promising evidence of the practicality of utilizing language model-based fuzz driver generation.}
The optimal configurations, namely <gpt-4-0613, ALL-ITER-K, 0.5>, achieved impressive success rates, effectively generating fuzz drivers that solved about 91\% (78/86) questions.
Moreover, three out of five LLMs assessed -- including an open-source option -- and half of the strategies explored can resolve over half of the questions.

\textbf{The substantial variation in success rates across different configurations underscores the significant influence of the three factors.}
By analyzing the data, we observe that results can greatly fluctuate when varying a single factor -- such as changing the temperature in a row, or switching models or prompting strategies in a column.
For example, <gpt-3.5-turbo-0613, NAIVE-1, 0.0> failed to solve any questions, whereas <gpt-3.5-turbo-0613, ALL-ITER-K, 0.0> managed to correctly address 76\% (65/86) of them. 
This indicates that achieving a high solve rate relies heavily on avoiding suboptimal combinations of factors.
Given that the table is sorted to reflect performance trends, the better outcomes tend to cluster in 'green areas', highlighting configurations where all contributing factors are well-adjusted.

\subsection{Analysis of Effectiveness Factors}

\noindent
\textbf{Prompt Strategies.} \tab 
The observed impacts of different prompting strategies exceeded our initial expectations during their design phase.
A comparison between NAIVE-1 and ALL-ITER-K showcases a dramatic improvement in optimal question solve rates, soaring from 10\% to 90\%, emphasizing the critical role of prompt design on tool effectiveness. 
To better understand the performance trends, Table~\ref{tab:overall_eva_rslt} presents the prompting strategies ranked by their overall effectiveness.
The trends in the results are intuitive: \textbf{in general, strategies that more comprehensively leverage available information tend to yield superior results}.
For example, the strategy UGCTX-K markedly outperforms BACTX-K.
This can be attributed to UGCTX-K's inclusion of example code snippets that illustrate certain usage of the target API.
A notable performance discrepancy is also seen when comparing BA-ITER-K with BACTX-K.
Despite starting with the same initial information, BA-ITER-K significantly surpasses BACTX-K.
The reason for this performance difference lies in BA-ITER-K's iterative method -- collecting debugging information to guide the model to fix the previous fuzz driver if it is ineffective.
Among all the strategies, ALL-ITER-K stands out as the most effective across different combinations of temperature settings and models.
This makes sense considering that ALL-ITER-K not only incorporates all extended API information but also adopts a recursive problem-solving methodology.
Conclusively, its design leads to the superior performance in our evaluation.
The detailed analysis of these strategies are discussed in Section~\ref{sec:rq3}.

\noindent
\textbf{Temperatures.} \tab 
Table~\ref{tab:overall_eva_rslt} clearly demonstrates that \textbf{configurations with a temperature setting of 0.5 tend to achieve the highest success rates}.
In contrast, models under a temperature setting above 1.0 experience a noticeable drop in performance.
Interestingly, it appears that \textbf{in general, lower temperatures, especially below the threshold of 1.0, show substantial performance advantage compared to models operating at higher temperatures}.
A surprising outcome is that models with 0.0 temperature perform remarkably well.
For instance, both <gpt-4-0613, ALL-ITER-K, 0.0> and <gpt-3.5-turbo-0613, BA-ITER-K, 0.0> stand out as second-best configuration when compared across the various temperature settings.
These results are reasonable considering the nature of fuzz driver generation task.
With a lower temperature setting, models tend to generate more consistent and predictable outputs, which benefits the synthesis of high-quality code.
High temperatures, while fostering creativity and randomness, may not provide any notable advantages in this context.
Specifically, these features are either substituted by the randomness contained in prompt strategies or deemed irrelevant by the assessment criteria.
For example, a prompting strategy like ALL-ITER-K inherently contains a built-in search process that brings the randomness from model input.
And the evaluation strictly assesses the quantity of effective drivers without considering the API usage diversity.
This criteria fits our evaluation goal, but discounts the creative diversity that could be introduced by higher temperatures.

\noindent
\textbf{Open-Source LLMs vs Closed-Source LLMs.} \tab 
As commonly understood in the industry, closed-source LLMs tend to outperform their open-source counterparts.
Among these, gpt-4-0613 is considered the front-runner in terms of generation capabilities.
Following closely behind is gpt-3.5-turbo-0613, which offers a cost-effective alternative due to its significantly lower token pricing.
However, it's worth noting that in the open-source domain, wizardcoder-15b-v1.0 has made remarkable strides, even surpassing Google's closed-source model, text-bison-001.
While wizardcoder-15b-v1.0 is nearly on par with gpt-3.5-turbo-0613, certain performance gaps can still be observed, but it stands as a commendable achievement for an open-source model.


\subsection{How Far Are We to Total Practicality?}
The above evaluation indicates that with the optimal configuration <gpt-4-0613, ALL-ITER-K, 0.5>, the LLM can solve 91\% predefined questions.
In other words, it can produce at least one effective fuzz driver for 78 out of 86 APIs examined.
However, this does not necessarily mean that LLMs are ready to be used in production.
Upon further examination of our APIs, we identified three primary challenges and detailed them as follows.

\begin{figure}[t]
    \centering
    \includegraphics[width=0.8\columnwidth]{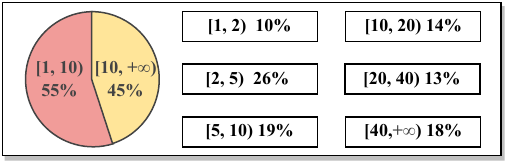}
    \caption{Question Cost Distribution For All Configurations on Resolved Questions.
    \footnotesize{
        \revision{
        $Cost\ of\ A\ Question = \frac{\#\ of\ Queries}{\#\ of\ Solutions}$.
        }
        }
    }
    \label{fig:topx-high-cost-pie}
\end{figure}


\noindent
\textbf{C1: High Token Cost in Fuzz Driver Generation.} \tab 
Though many configurations have shown a high rate of successful problem resolution, our analysis indicates the results come with substantial costs.
The data in Figure~\ref{fig:topx-high-cost-pie} details the percentage of high cost questions for all evaluated configurations.
Remarkably, it reveals that on average, resolving 45\% questions entail costs exceeding 10.
This suggests that for 50\% of the resolved questions, a prompt-based strategy may yield just one effective fuzz driver after repeating the entire query process 10 times or more.
When considering only the questions with costs surpassing 20 or even 40, the percentages remain notable at 31\% and 18\%, respectively.
These findings underscore a strong incentive for further research into cost reduction techniques.
Reducing costs is not only a practical concern with direct financial consequences but also essential for improving the efficiency of LLM-based fuzz driver generation.

\noindent
\textbf{C2: Ensuring Semantic Correctness of API Usage.} \tab 
In our evaluation, we found that there is a discrepancy for approximately 34\% (29/86) of the APIs -- assuming LLMs can successfully create at least one effective fuzz driver for each in the evaluation setting, this success cannot be translated into the full automation of fuzz driver generation for them.
The issue at hand lies in the potential misuse of APIs within the generated drivers, which requires validation to ensure semantic correctness.
For example, LLMs may incorrectly initializing the argument of an API, such as passing a mutated filename to the API instead of passing a created file first and then mutating its content for fuzzing, or missing some condition checks before calling API.
In our evaluation process, we manually implemented semantic checkers to identify such API misuses for accurate assessment (details on our semantic checkers are provided in Section~\ref{sec:eval-framework}). 
However, fully automating the validation of semantic correctness remains a significant hurdle.
Consequently, even though it is feasible to generate effective fuzz drivers with the help of these LLMs, distinguishing them from the ineffective ones can be problematic due to the absence of automated methods for validating semantic correctness.
This challenge underscores the need for developing robust techniques to automatically ensure the semantic accuracy of generated fuzz drivers before they can be reliably deployed in production.


\textbf{C3: Satisfying Complex API Usage Dependencies.} \tab 
Overall, there are five questions cannot be resolved by any assessed configurations.
These questions are challengeable since their driver generation requires the deep understanding of specific contexts.
For instance, generating the driver for \texttt{tmux}~\cite{tmux-ossfuzz-driver-link} requires the construction of various concepts, such as session, window, pane, etc, and their relationships.
Similarly, for network-related questions~\cite{libmodbus-ossfuzz-driver-link, civetweb-ossfuzz-driver-link}, a standby network server or client is required to be created before calling the target API.
The effective drivers can only be generated by respecting these specific contextual requirements.

\noindent
\begin{tcolorbox}[size=title, opacityfill=0.1, breakable]
While LLM-based generation has shown promising potential, it still faces certain challenges towards high practicality.
\end{tcolorbox}

\section{Fundamental Challenge (\textbf{RQ2})}




\subsection{Links Between Question and Performance}

\begin{figure}[t]
\centering
\includegraphics[width=0.8\linewidth]{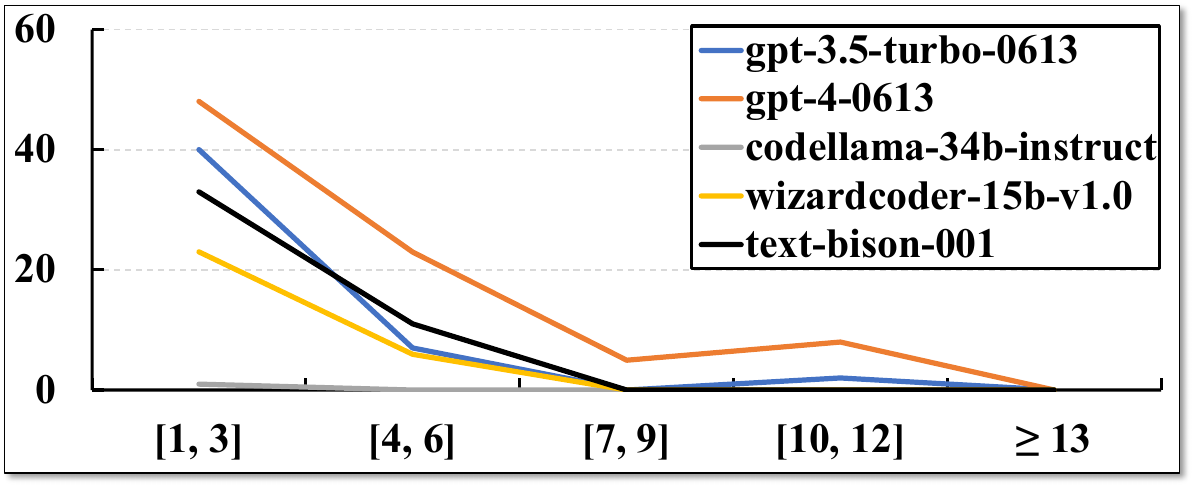}
\caption{Average Query Success Rate Per Question Complexity Score Bucket.
\footnotesize{
Data: All BACTX-K (K = 40) configurations.
}
}
\label{fig:rq2-query-succ-rates-per-score}
\end{figure}


To investigate the core difficulties in generating fuzz drivers with LLMs, we scrutinized the outcomes of the BACTX-K strategy.
\revision{
This strategy is a proper starting point for understanding the fundamental challenges since it merely uses generally accessible information and has simple query workflow.
}
In Figure~\ref{fig:rq2-query-succ-rates-per-score}, \textbf{there is a clear inverse proportion relationship between the query success rate and the complexity of a question, irrespective of the used models and temperatures}.
The complexity of a question is measured by first constructing the minimal fuzz driver of each question and then quantifying the API specific usage contained in the minimized code.
A minimal effective driver for a question is created based on the OSS-Fuzz driver by removing the unnecessary part of the code and replacing the argument initialization into a simpler solution according to the cases enumerated in Section~\ref{sec:preliminaries}.
Then the complexity is quantified as the sum of the count of the following elements inside code:
\ding{182} unique project APIs;
\ding{183} unique common API usage patterns;
\ding{184} unique identifiers including non-zero literals and project global variables excluding the common API usage code;
\ding{185} branches and loops excluding the common API usage code.
Note that all branches of one condition will be counted as one.
Overall, \ding{182}, \ding{184} measure API specific vocabularies while \ding{185} for API specific control flow dependencies'.
We put detailed calculation examples at \cite{fuzz-drvier-study-website}.
\compactline

Considering the generation process, it is intuitive that \textbf{LLMs' performance degrades when the complexity of target API specific usage increases}.
To generate effective drivers, LLMs should at least generate code satisfying minimal requirements.
In other words, they must accurately predict the API argument usage and control flow dependencies.
However, this is challenging since LLMs cannot validate their predictions against documentation or implementations as humans do.
It is reasonable to assume that LLMs have learned the language basics and common programming practices due to their training on vast amounts of code.
But the API specific usage, such as the semantic constraints on the argument, cannot be assumed.
On one hand, there may only have limited data about this in training.
On the other hand, details can be lost during preprocessing or the learning stage while the accurate generation is required.
Therefore, the more API usage a LLM needs to predict, the greater the likelihood of errors, particularly for less common usages that do not follow the mainstream design patterns or have special semantic constraints.
Such situations are common in C projects, whose APIs often contain low-level project-specific details.


\noindent
\begin{tcolorbox}[size=title, opacityfill=0.1, breakable]
The performance of LLM-based generation declines significantly when the complexity of API specific usage increases.
\end{tcolorbox}

\begin{figure}[t]
\centering
\includegraphics[width=1.0\linewidth]{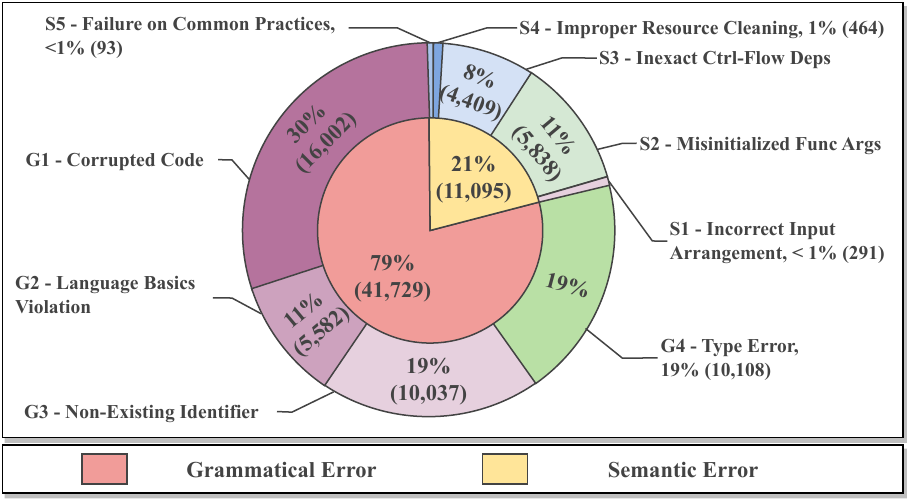}
\caption{Failure Taxonomy.
\footnotesize{
Data: BACTX-K (K = 40) configurations.
}
}
\label{fig:rq2-failure-taxonomy}
\end{figure}

\subsection{Failure Analysis}
\label{sec:failure-analysis}

%
%
%
%
%

To understand how the generation fails on API specifics, we conducted failure analysis on \texttt{BACTX-K}.
The direct failure reason of the driver is collected to reveal the generation blockers.
In total, 52,824 ineffective drivers were analyzed.
The runtime errors of 11,095 drivers are semi-automatically analyzed while the compilation and link errors are categorized based on the compiler outputs.

\noindent
\textbf{Failure Taxonomies.} \tab 
Figure~\ref{fig:rq2-failure-taxonomy} details the root cause taxonomy.
There are nine root causes fallen into two categories:
the grammatical errors reported by compilers in build stage, and the semantic errors which are abnormal runtime behaviors identified from the short-term fuzzing results.
\ding{182} \textit{G1 - Corrupted Code}, the drivers do not contain a complete function of code due to either the token limitation or mismatched brackets;
\ding{183} \textit{G2 - Language Basics Violation}, the code violates the language basics like variable redefinition, parentheses mismatch, incomplete expressions, etc;
\ding{184} \textit{G3 - Non-Existing Identifier}, the code refers to non-existing things such as header files, macros, global variables, members of a \texttt{struct}, etc;
\ding{185} \textit{G4 - Type Error}. One main subcategory here is the code passes mismatched number of arguments to a function.
The rest are either unsupported type conversions or operations such as calling non-callable object, assigning \texttt{void} to a variable, allocating an incomplete \texttt{struct}, etc;
\ding{186} \textit{S1 - Incorrect Input Arrangement}, the input size check either is missed when required or contains an incorrect condition;
\ding{187} \textit{S2 - Misinitialized Function Args}, the value or inner status of initialized argument does not fit the requirements of callee function.
Typical cases are closing a file handle before passing it, using wrong enumeration value as option parameter, missing required APIs for proper initialization, etc;
\ding{188} \textit{S3 - Inexact Ctrl-Flow Deps}, the control-flow dependencies of a function does not properly implemented.
Typical cases are missing condition checks such as ensuring a pointer is not \texttt{NULL}, missing APIs for setting up execution context, missing APIs for ignoring project internal abort, using incorrect conditions, etc.
\ding{189} \textit{S4 - Improper Resource Cleaning}, the cleaning API such as \texttt{xxxfree} is either missing when required or is used without proper condition checks; 
\ding{190} \textit{S5 - Failure on Common Practices}, the code failed on standard libraries function usage like messing up memory boundary in \texttt{memcpy}, passing read-only buffer to \texttt{mkstemp}, etc.
Examples of these categories are shown in ~\cite{fuzz-drvier-study-website}.

Overall, the failures cover API usages in various dimensions: from grammatical level detail to semantic level direction, and from target API control flow conditions to dependent APIs' declarations.
Improving this is challengeable since:
\ding{182} \textbf{the involved usage is too broad to be fully put into one prompt}, which may either exceed the token limitation or distract the model;
\ding{183} \textbf{the useful usage for generating one driver cannot be fully predetermined}.
On one hand, models are inherently blackbox and probabilistics, whose mistakes cannot be fully predicted.
On the other hand, there are usually multiple implementation choices for a given API.





\noindent
\begin{tcolorbox}[size=title, opacityfill=0.1, breakable]
\revision{
Most failures are of mistakes in API usage specifics.
The broadness of the involved usage is the major challenge.
}
\end{tcolorbox}
\section{Characteristics of Key Design (\textbf{RQ3})}

\begin{figure}[t]
\centering
\begin{subfigure}[b]{0.48\linewidth}
\centering
\includegraphics[width=\textwidth]{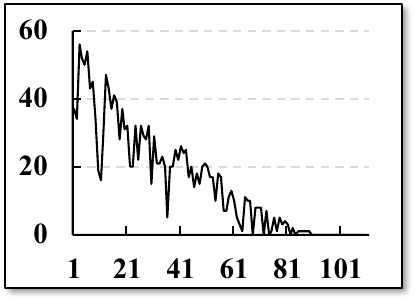}
\caption{
\footnotesize{Number of Questions Solved by Repeat for All Configurations.}
}
\label{fig:eff-repeatedly-query-a}
\end{subfigure}\hfill
\begin{subfigure}[b]{0.48\linewidth}
\centering
\includegraphics[width=\textwidth]{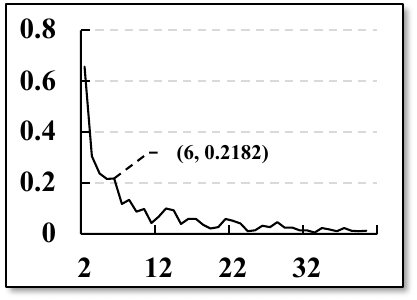}
\caption{
\footnotesize{Avg Pct of Solved Questions Per Repeat Round in Top-20 Configs.}
}
\label{fig:eff-repeatedly-query-b}
\end{subfigure}

\caption{ 
Statistics on the Effectiveness of Repeatedly Query.
}
\label{fig:eff-repeatedly-query}
\end{figure}

\subsection{Repeatedly Query}
Repeated querying is a critical aspect of prompt strategies, greatly enhancing the success rate in generating fuzz drivers regardless of employed models, temperatures, and prompt designs.
Specifically, for the optimal configuration <gpt-4-0613, 0.5, ALL-ITER-K>, approximately 47.44\% of the issues were resolved by reinitiating the query process (37 out of 78 total resolved issues were solved upon repetition).
For the top-20 configurations, this contribution remains significantly high at an average of 67.50\%.

Figure~\ref{fig:eff-repeatedly-query-a} displays the count of questions resolved through repeated querying across all evaluated configurations, ranked by their overall effectiveness as detailed in Table~\ref{tab:overall_eva_rslt}.
This demonstrates a direct correlation between the benefit of repeated queries and the efficacy of the configuration—\textbf{the more effective a configuration, the greater the gains from repeating the queries}.

Additionally, Figure~\ref{fig:eff-repeatedly-query-b} presents the average percentage of questions resolved in each subsequent round of querying for the top-20 configurations.
Here, the percentage for round X is determined by $\frac{Rslt(X) - Rslt(X - 1)}{Rslt(1)}$, with $Rslt(X)$ indicating the number of questions resolved by round X.
The X-axis starting from round two, highlighting that the first round corresponds to the initial query.
This data shows that \textbf{the gain of repeated queries drops significantly after the initial few rounds}.
From our evaluation, we recommend limiting repeated queries to no more than six, where the sixth round still manages to resolve an additional 20\% of questions compared to the results of the first round.

\begin{figure}[t]
\centering
\begin{subfigure}[b]{0.48\linewidth}
\centering
\includegraphics[width=\textwidth]{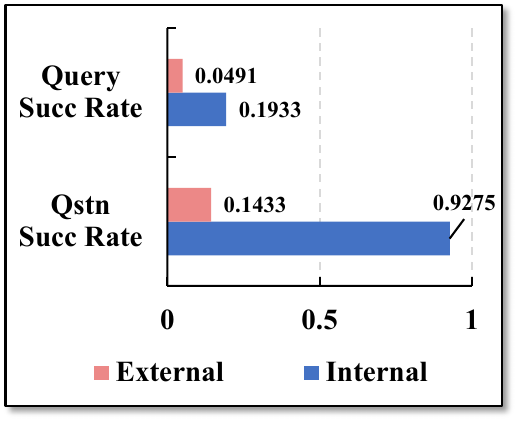}
\end{subfigure}\hfill
\begin{subfigure}[b]{0.48\linewidth}
\centering
\includegraphics[width=\textwidth]{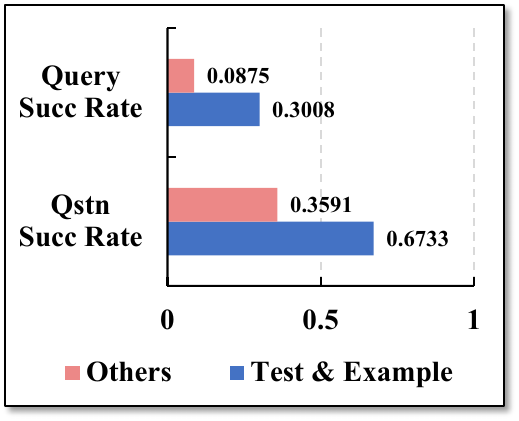}
\end{subfigure}

\caption{Average Query/Question Success Rate of Different Example Sources for All Configurations.}
\label{fig:succ-rate-of-different-ex-sources}
\end{figure}

\subsection{Query With Extended Information}

\noindent
\textbf{Querying With API Documentation.}
\tab
By comparing DOCTX-K and BACTX-K, we found that \textbf{there is no significant changes between their results in the metrics of resolved questions}.
On one hand, a significant percentage (43\%) of APIs in the evaluated questions do not have API documentation (49 out of 86 have).
When there is no documentation for an API, the DOCTX-K queries are identical to BACTX-K's.
On the other hand, adding API documentation in the queries may not provide enough details directly stating the API usage.
This is because these API documentations usually contain a high-level description of the usage, typically a summary of main functionality with one-sentence explanations for arguments. 
However, the blocker-solving usage information discussed in Section~\ref{sec:failure-analysis}, such as low level argument initialization specifics, control flow dependencies, or the usages of its dependent APIs, is usually not included.

\noindent
\begin{tcolorbox}[size=title, opacityfill=0.1, breakable]
API documentation has minor performance benefits due to the limited usage description it contained.
\end{tcolorbox}


\noindent
\textbf{Querying With Example Code Snippets.} 
\tab
When comparing the results of BACTX-K and UGCTX-K presented in Table~\ref{tab:overall_eva_rslt}, we can clearly observe that incorporating example code snippets substantially enhances performance in most configurations.
In particular, the addition of example snippets results in an average resolution of 104\% more questions across the 22 evaluated configurations, which includes five models and five different temperature settings.

Nonetheless, further analysis reveals that \textbf{the inclusion of usage examples incurs a much higher token cost, with an average increase of tenfold}.
The ratio of token costs for these two approaches varies from 4.20 to 39.71 across all configurations, with an average ratio of 14.65.
Notably, the UGCTX-K approach demands an average of 32,367 tokens to generate a single correct solution.

Figure~\ref{fig:succ-rate-of-different-ex-sources} depicts our investigation into the impact of different sources of example snippets on the quality of solutions.
This figure assesses the success rates of queries/questions associated with various example sources, which are categorized in two distinct manners based on their file paths: first, as \ding{182} \textit{External} vs. \textit{Internal}, with \textit{Internal} comprising the target project and its variations, and \textit{External} consisting of all other sources; second, as \ding{183} \textit{Test \& Example} vs. \textit{Others}, where the first group includes files with paths that contain "test" or "example" in any capitalization.
The underlying data for these plots stems from questions that were solved by UGCTX-K but not by BACTX-K across all tested configurations.
According to this analysis, it is clear that \textbf{both \textit{Internal} and \textit{Test \& Example} sources are associated with significantly higher quality example snippets in comparison to their counterparts}.

\noindent
\textbf{Case Studies.}
\tab
\# 9 \texttt{wc\_Str\_conv\_with\_detect}
\tab
This case is challenging due to the unintuitiveness of its API usage.
The API declaration is "\texttt{Str wc\_Str\_conv\_with\_detect(Str is,wc\_ces * f\_ces,wc\_ces hint,wc\_ces t\_ces)}".
It is used for converting the input stream \texttt{is} from one CES (character encoding scheme, \texttt{f\_ces}) to another (\texttt{t\_ces}).
Most basic strategy drivers made mistakes on the creations of either \texttt{is} (the confusing type \texttt{Str}) or CESs, where \texttt{is} has to be created using particular APIs like \texttt{Strnew\_charp\_n} and CESs should be specific macros or carefully initialized \texttt{struct}.
Example helps here by directly providing the usage to models.

\# 37 \texttt{igraph\_read\_graph\_graphdb}
\tab
The hardest part in this case is the implicit control flow dependency it required.
Besides correctly initializing the arguments, it has to call an API to mute the builtin error handlers.
By default, the API will abort immediately when any abnormal input is detected, which causes frequent false crashes blocking the fuzzing progress.
To mute it, the driver needs to custmoize the error handler, \textit{e.g.}, call \texttt{igraph\_set\_error\_handler\-(igraph\-\_error\_handler\_ignore)}.
This requirement is hard to be inferred beforehand due to its semantic nature and few inference clues.
However, some unit tests in project such as \texttt{foreign\_empty.c} contain this usage, which directly instructs the generation.

\begin{figure*}[t]
\centering
\begin{subfigure}[t]{\linewidth}
\centering
\includegraphics[width=0.495\textwidth]{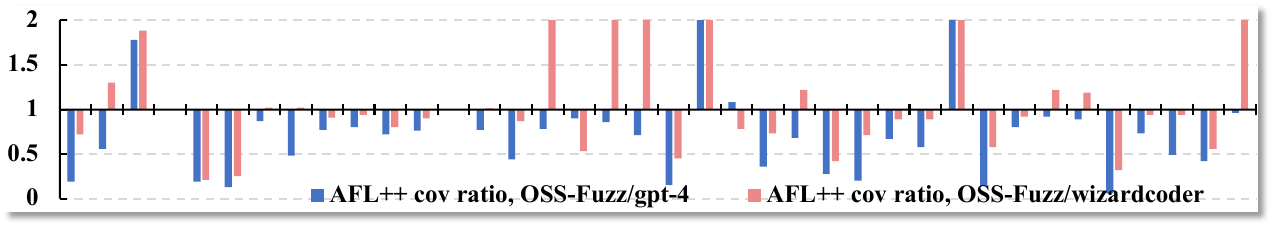}
\hfill
\includegraphics[width=0.495\textwidth]{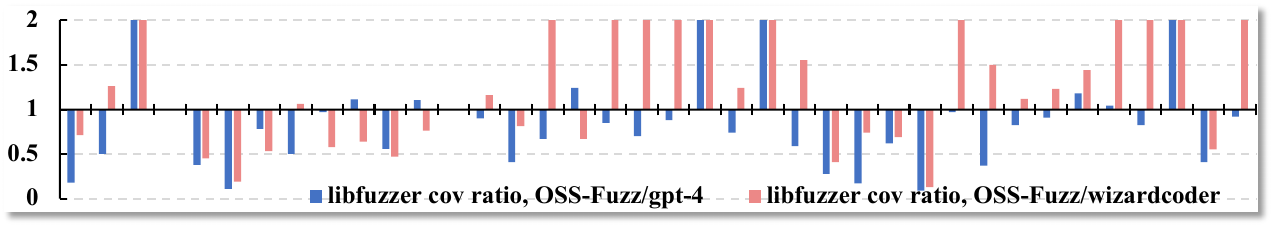}
\vspace{-3pt}
\caption{Comparison in Average Coverage Ratio.
\footnotesize{
Ratio = OSS-Fuzz/LLM,
ratio < 1 $\rightarrow$ OSS-Fuzz's driver has lower coverage.
}
}
\label{fig:cmp-cov}
\end{subfigure}%
\\
\begin{subfigure}[t]{\linewidth}
\centering
\includegraphics[width=0.495\textwidth]{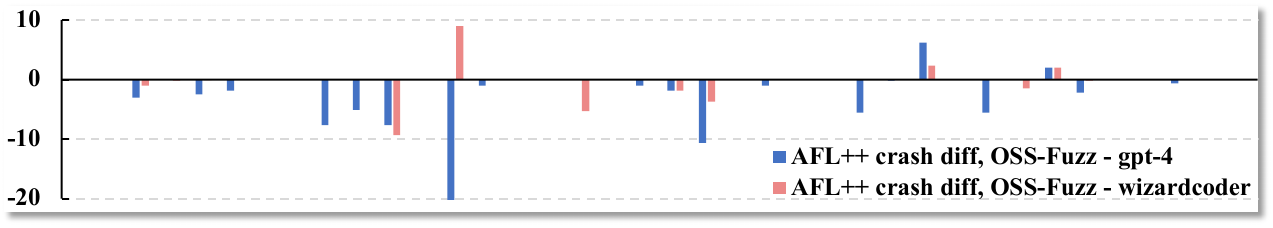}
\hfill
\includegraphics[width=0.495\textwidth]{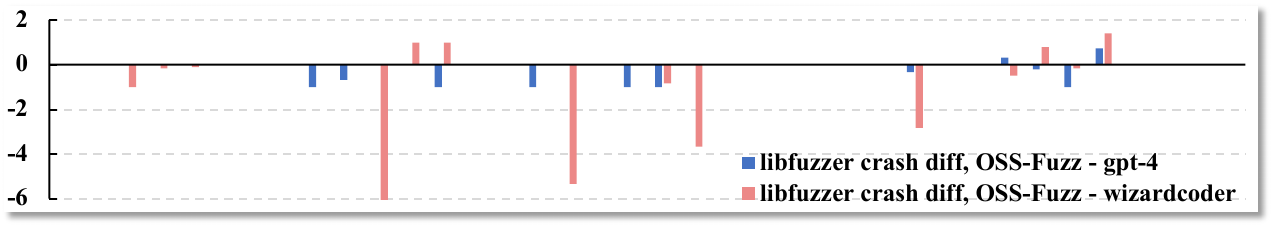}
\caption{Comparison in the Difference of Average Unique Crashes.
\footnotesize{
Diff = OSS-Fuzz - LLM,
diff < 0 $\rightarrow$ OSS-Fuzz's driver find less unique crashes.
}
}
\label{fig:cmp-crash}
\end{subfigure}%
\caption{Metric Comparison of LLM-Generated and OSS-Fuzz Drivers.
\footnotesize{
Y-axis is the metric.
For clarity, question id (x-axis) is omitted.
}
}
\label{fig:rq4-cmps}
\end{figure*}
\noindent
\begin{tcolorbox}[size=title, opacityfill=0.1, breakable]
    \revision{
Example code snippets can greatly enhance model performance by providing direct insights on API usage.
    }
"test/example files", "code files from the target/variant projects" are high quality sources.
\end{tcolorbox}

\subsection{Iterative Query}
The iterative query strategy is another key design that can lead to significant improvements in performance.
Referring to Table~\ref{tab:overall_eva_rslt}, we find that, on average, incorporating an iterative query strategy into BACTX-K -- that is, adopting the BA-ITER-K approach -- helps solve 159\% more questions.
Similarly, ALL-ITER-K resolves 23\% more questions than UGCTX-K. 
However, this strategy does come at a cost.
\textbf{The inclusion of iterative design tends to lead to higher token usage when generating correct solutions}.
On average, the iterative strategy increases token costs by 57\% for BACTX-K per successful driver generation and by 17\% for UGCTX-K.

The effectiveness of the iterative strategy can be attributed to two key factors.
Firstly, \textbf{it leverages a wider array of information}, including error data generated from validating previously generated drivers.
Secondly, \textbf{it tackles the problem incrementally}, employing a step-by-step, divide-and-conquer approach that simplifies the complexity of the generation task.
This methodology is exemplified in the case studies that follow, illustrating how the iterative strategy typically operates through practical examples.

\noindent
\textbf{Case Studies.}
\tab
\#5 \texttt{md\_html}
\tab
This API requires the preparation of a customized callback function pointer as the argument, where all previous strategies failed to figure out.
The callback function is used to handle the output data of API.
All the drivers generated by \texttt{UGCTX} either pass a \texttt{NULL} pointer or a non-existing function name.
Iterative query guides the fix by providing the link error highlighting that this referred function is undefined.

\#73 \texttt{pj\_stun\_msg\_decode}
\tab
This is another typical case why iterative strategy works.
The initialization of its first argument has multi-level API dependencies.
The dependency chain is: 
\ding{182} the API ->
\ding{183} \texttt{pj\_pool\-\_create} -> 
\ding{184} \texttt{pj\_caching\_pool\-\_init}, where -> means depends.
All non-iterative strategies failed to prepare a driver with all correct usage detail of these indirect dependencies while iterative strategies solve this by providing error related feedback to LLMs and solving multiple errors one by one.
In one of the solved iterative query, it first corrects the incorrect used API of \ding{184}, then figures out the mismatched type error when calling \ding{183}.
Lastly, for the driver's runtime crash, LLMs use two rounds to fix according to the assertion code located from crash stacks.


\noindent
\begin{tcolorbox}[size=title, opacityfill=0.1, breakable]
    \revision{
Iterative query helps in utilizing more diverse information and solving the problem in a step-by-step manner.
However, it has higher token cost and increased complexity.
    }
\end{tcolorbox}

\section{OSS-Fuzz Driver Comparison (\textbf{RQ4})}
\label{sec:rq3}

\noindent
\textbf{Comparison Overview.}
\tab
We compared LLM-generated drivers with OSS-Fuzz's to obtain more practical insights.
Note that OSS-Fuzz drivers are practically used in industry for continuous fuzzing and most of them are manually written and improved for years.
Particularly, LLM-generated drivers under comparison are from gpt-4-0613 and wizardcoder-15b-v1.0 using iterative strategies with temperature 0.5.
These two configurations are the best representative for closed-source and open-source LLMs.
In total, we evaluated 53 questions which are both resolved by all configurations.
Multiple drivers of one question are merged as one to ease the comparison.
This is done by adding a wrapper snippet which links the seed scheduling with the selection of the executed logic from merged drivers.
Specifically, a switch structure is added to determine which driver it will execute based on a part of the input data.
During each fuzzing iteration, only the logic of one merged driver is executed.
Besides, some compound OSS-Fuzz drivers are designed to fuzz multiple APIs.
For clear comparison, we merged all drivers of questions involved in one compound driver as one.
In total, we prepared 38 drivers for each assessed LLM or OSS-Fuzz.
The comparisons cover both code and fuzzing metrics such as the number of used APIs, oracles, coverage, and crashes.

\noindent
\textbf{Fuzzing Setup.}
\tab
Considering the randomness of fuzzing, we followed the suggestions from ~\cite{fuzz-eval}: 
the fuzzing experiments are conducted with five times of repeat for collecting average coverage information and the fuzzing of each driver lasts for 24 hours.
We used \texttt{libfuzzer}~\cite{libfuzzer} and \texttt{AFL++}~\cite{fioraldi2020afl++} as fuzzers with empty initial seed and dictionary.
"\texttt{-close\_fd\_mask=3 -rss\_limit\_mb=2048 -timeout=30}" is used for \texttt{libfuzzer} while \texttt{AFL++}'s is the default setup of \texttt{aflpp\_driver}.
For fair comparison, the coverage of fuzz driver itself is excluded in post-fuzzing data collection stage (the merged driver can have thousands of lines of code) but kept in fuzzing stage for obtaining coverage feedback.
In total, the experiments took 3.75 CPU year.




\noindent
\textbf{Code Metric: API Usage.}
\tab
The API usage is measured by the number of unique project APIs used in the fuzz driver.
Overall, 14\% (17/35) \texttt{gpt-4-0613} drivers have used less project APIs than OSS-Fuzz's while 39\% for \texttt{wizardcoder-15b-v1.0}.
By manually investigating these drivers, we found that \textbf{LLMs conservatively use APIs in driver generation if no explicit guidance in prompts}.
For instance, some drivers only contain necessary usages such as argument initialization.
And the API usage is hardly extended such as adding APIs to use an object after parsing it.
This is a reasonable strategy since aggressively extending APIs increases the risk of generating invalid drivers.
Adding example snippets in the prompt can alleviate this situation.
As for OSS-Fuzz drivers, the API usage diversity is case by case since they are from different contributors.
Some drivers, \textit{e.g.},~\cite{croaring-ossfuzz-driver-link} are minimally composed and some are extensively exploring more features of the target, \textit{e.g.},~\cite{lua-ossfuzz-driver-link}.
One interesting finding is that some OSS-Fuzz drivers are modified from the test files rather than written from scratch, which is a quite similar process as querying LLM with examples.
For example, \texttt{kamailio} driver~\cite{kamailio-ossfuzz-driver-link} is modified from test file~\cite{kamailio-test-example-link}.
Prompting with this example, LLM can generate similar driver code.


%

\noindent
\textbf{Code Metric: Oracle.}
\tab
We did statistics on the oracles of the drivers.
The result is quite clear:
in all 78 questions resolved by LLMs, OSS-Fuzz drivers of 15 questions contain at least one oracle which can detect semantic bugs, while there are \textbf{no LLM-generated drivers have oracles}.
The used semantic oracles can be categorized as following:
\ding{182} check whether the return value or output content of an API is expected, \textit{e.g.}, ~\cite{bind9-api-output-oracle};
\ding{183} check whether the project internal status has expected value, \textit{e.g.}, ~\cite{igraph-internal-status-oracle};
\ding{184} compare whether the outputs of multiple APIs conform to specific relationships, \textit{e.g.}, ~\cite{bind9-check-two-apis-oracle}.


\noindent
\begin{tcolorbox}[size=title, opacityfill=0.1, breakable]
LLMs tend to generate fuzz drivers with minimal API usages, significant space are left for further improvement such as extending the use of API outputs or adding semantic oracles.
\end{tcolorbox}


\noindent
\textbf{Fuzzing Metric: Coverage and Crash.}
\tab
Figure~\ref{fig:cmp-cov},~\ref{fig:cmp-crash} plot the coverage and crash comparison results.
Instead of presenting every detail of the experiments for hundreds of drivers, the plots lists the comparison in certain metrics while the full experiment details can be found at~\cite{fuzz-drvier-study-website}.
Overall, \textbf{in most questions, the LLM-generated drivers demonstrate similar or better performance in metrics of both coverage and the number of uniquely found crashes}.
Note that there are no false positive since the generated fuzz drivers are already filtered by the semantic checkers provided from our evaluation framework.
If only the fully automatic validation process are adopted, \textit{i.e.}, removing the last two checkers in Figure~\ref{fig:validation-checker}, the fuzzing outcome will be messed with huge number of false positives, incurring significant manual analysis efforts.

\noindent
\begin{tcolorbox}[size=title, opacityfill=0.1, breakable]
LLM-generated drivers can produce comparable fuzzing outcomes as OSS-Fuzz drivers.
In large scale application, how to practically pick effective fuzz drivers is the major challenge.
\end{tcolorbox}

\section{Discussion}
\label{sec:ttv}

\revision{
\noindent
\textbf{Relationships With OSS-Fuzz-Gen.}
\tab
The Google OSS-Fuzz team has undertaken a parallel work called OSS-Fuzz-Gen~\cite{oss-fuzz-gen} for LLM-based fuzz driver generation.
Their public information contains one security blog~\cite{oss-fuzz-gen-blog} and the source code repository~\cite{oss-fuzz-gen}.
Our work is complementary to theirs.
Overall, at the time of submission, they put high efforts on filling the engineering gap between LLM interfaces and OSS-Fuzz projects.
Their experiments are conducted on top commercial LLMs, aiming to showcase that LLM-generated fuzz drivers can help in finding zero-day vulnerabilities and reaching new testing coverage.
However, there few discussion on the fundamental questions such as the design choices behind their prompt strategy, the pros and cons for different strategies, how the effectiveness varies for different models and parameters, and what are the inherent challenges and potential future directions.
Our study, on the other hand, complements theirs by exploring these fundamental issues.
We carefully designed prompt strategies, evaluated them on various models (open and commerical LLMs) and temperatures, and distilled findings from the results. 
}

\revision{
\noindent
\textbf{Contributing to OSS-Fuzz-Gen.}
\tab
We carefully examined the prompt strategies of OSS-Fuzz-Gen from their implementation and validated where our insights can help.
Interestingly, their current strategy support part of our insights.
For instance, they adopted 10 time repeat results~\cite{oss-exp-repeat} and used a lower temperature (0.4) in experiments~\cite{oss-default-temperature}.
Besides, we found that OSS-Fuzz-Gen only identifies and fixes build errors while ignoring the runtime errors caused by driver.
Their generation ends when a compilable fuzz driver is synthesized and then they manually checks the validity of these drivers.
To improve this, we implemented our strategies for drivers with fuzzing runtime errors in their platform, including the identification (automatic part of validation process)~\cite{oss-pr-191, oss-pr-187, oss-pr-199, oss-pr-185}, categorization, and the corresponding iterative fix procedure~\cite{oss-pr-204, oss-pr-198}.
These enhancements added new functionalities refining the generation results, where the cases showing its effectiveness are quickly identified~\cite{oss-pr-effective-case, oss-pr-fp-filter-case} during their benchmark tests (29 APIs, 18 projects).
Currently, the improvement is merged into the main branch and is actively used to fuzz all 282 supported projects, marking a significant milestone to us.
We are keeping refine our commitments, such as integrating more fine-grained error information during fix. 
}

\revision{
\noindent
\textbf{Potential Improvements.}
\tab
From our perspective, to improve the performance of LLM-based fuzz driver generation, efforts from three dimensions can be further explored.
First, the domain knowledge contained inside the target scope can be modeled and utilized for better generation.
For instance, to test network protocol APIs, the communication state machine of that protocol can be learned first and then used to guide the driver generation.
Besides, more sophisticated prompt-based solutions can be explored, such as hybrid approaches combining traditional program analysis and prompt strategies, or agent-based approaches.
Lastly, fine-tuning based methods is also a promising direction since this can enhance both the generation's effectiveness and efficiency from a model level.
}

\noindent
\textbf{Threat to Validity.}
\tab
One internal threat comes from the effectiveness validation of the generated drivers.
To address this, we carefully examined the APIs and manually wrote tests for them to check whether the semantic constraints of a specific API have been satisfied or not.
Another threat to validity comes from the fact that some OSS-Fuzz drivers, \textit{e.g.}, code written before Sep 2021, may already be contained in the model training data, which raises a question that whether the driver is directly memorized by the model from the training data.
Though it is infeasible to thoroughly prove its generation ability, which requires the retrain of LLMs, we found several evidences that supports the answers provided by these models are not memorized:
Many generated drivers contain APIs that do not appear in the OSS-Fuzz drivers, especially for those drivers hinted by example usage snippets or iteratively fixed by usage and error information.
Besides, the generated drivers share a distinct coding style as OSS-Fuzz drivers.
For example, the generated code are commented with explanation on why the API is used and what it is used for, etc.
The main external threat to validity comes from our evaluation datasets.
Our study focused on C projects while the insights may not be necessarily generalizable to other languages.
\compactline



\section{Related Work}

\noindent
\textbf{Fuzz Driver Generation.} \tab 
Several works~\cite{fuzzgen, fudge, apicraft, winnie, intelligen, rubick, utopia, chen2023hopper} have focused on developing automatic approaches to generate fuzz drivers.
Most of these works follow a common methodology, which involves generating fuzz drivers based on the API usage existed in consumer programs, \textit{i.e.,} programs containing code that uses these APIs.
For instance, by abstracting the API usage as specific models such as trees~\cite{apicraft}, graphs~\cite{fuzzgen}, and automatons~\cite{rubick}, several works propose program analysis-based methods to learn the usage models from consumer programs and conduct model-based driver synthesis.
In addition, a recent work~\cite{utopia} emphasizes that unit tests are high quality consumer programs and proposes techniques to convert existing unit tests to fuzz drivers.
Though these approaches can produce effective fuzz drivers, their heavy requirements on the quality of the consumer programs, \textit{i.e.,} the consumers must contain complete API usage and are statically/dynamically analyzable, limit their generality.
Furthermore, synthesized code often lacks human readability and maintainability, limiting their practical application.
\revision{
Some parallel works~\cite{oss-fuzz-gen,prompt-fuzz} have also explored the LLM-based fuzz driver generation.
However, their main goal is to build tools demonstrating the potential of LLM-based generation.
Our study complements them by focusing on delivering the first comprehensive understanding of the fundamental issues in this direction.
}
\compactline


\noindent
\textbf{LLM for Generative Tasks.} \tab 
Recent works have explored the potential of LLM models for various generative tasks, such as code completion~\cite{wei2023magicoder}, test case generation~\cite{guifill, zhanglingming-llm-are-zero-shot-fuzzers, yang2023white, deng2023large, schafer2023adaptive, siddiq2023exploring, xia2023universal} and code repairing~\cite{sp-repair, abhik-repair, xia2023keep}.
These works utilize the natural language processing capabilities of LLM models and employ specific prompt designs to achieve their respective tasks.
To further improve the models' performance, some works incorporate iterative/conversational strategies or use fine-tuning/in-context learning techniques.
In test case generation, previous research works have primarily targeted on testing deep learning libraries~\cite{zhanglingming-llm-are-zero-shot-fuzzers, deng2023large} and unit test generation~\cite{schafer2023adaptive, siddiq2023exploring}. 
Considering the intrinsic differences between fuzz drivers and other tests and the difference on studied programming languages, these works cannot answer the fundamental effectiveness issues of LLM-based fuzz driver generation, indicating the unique values of our study.

% ...

\vspace{-2pt}
\section{Conclusion}
\revision{
Our study centers around answering fundamental issues of LLM-based fuzz driver generation's effectiveness.
To do that, we designed a dataset and six prompt strategies, and did extensive evaluation on different models and temperatures.
Our study not only established the basic understanding on this direction but also indicates the potential future improvements.
Furthermore, our insights have been applied into industrial practical fuzz driver generation platform.
}
\compactline


\section{Data Availability}

The source code and data involved in our study can be found at~\cite{fuzz-drvier-study-website}.







\bibliographystyle{ACM-Reference-Format}
\bibliography{reference}



\end{document}